\documentstyle[prc,aps,epsf]{revtex}
\begin{document}
\title{The Pauli principle in a three-body cluster model and  
the momentum distributions after fragmentation of $^6$He and $^{11}$Li }

\author{E.~Garrido, D.V.~Fedorov and A.S.~Jensen \\
Institute of Physics and Astronomy,
Aarhus University, \\ DK-8000 Aarhus C, Denmark}
\date{\today}

\maketitle

\begin{abstract}
We investigate two simple prescriptions to account for the Pauli
principle in a three-body cluster model employing a new method based on
an adiabatic hyperspherical expansion to solve the Faddeev equations in
coordinate space. The resulting wave functions are computed and
compared. They are furthermore tested on halo nuclei by calculations of
momentum distributions and invariant mass spectra arising after
fragmentation of fast $^6$He and $^{11}$Li in collisions with light
targets. The prescriptions are very accurate and the available measured
quantities are remarkably well reproduced when final state interactions
are included.
\end{abstract}
\vspace{5mm}
\par\leavevmode\hbox {\it PACS:\ } 25.60.-t, 21.45.+v, 21.60.Gx
\vspace{5mm}

\section{Introduction} 

Three-body models are useful to describe the relative wave function of
three particles when their intrinsic structure remains unchanged
\cite{ric92,zhu93,fed94a}. The intrinsic and the relative degrees of
freedom are then assumed to be completely decoupled. However, when the
particles themselves are composite structures (clusters) and contain
identical fermions this assumption violates the Pauli principle related to
fermions distributed in different clusters. This violation typically
leads to unphysical deeply bound two-cluster states. In the two-body
problem this violation can be easily cured by removing this unphysical
state from the active space available for the system.

In the three-cluster problem most of the usual approaches to include
the Pauli principle basically fall into two categories: i) to project
out the undesired overlap of the three-body wave function with the Pauli
forbidden two-body state\cite{kuk78,mar82,ban83} and ii) to modify
the two-body interaction in such a way that the unphysical bound
cluster-cluster state is avoided, provided the low energy scattering
properties \cite{joh90} or the phase shifts \cite{bay87,fie90} remain
unchanged.  The prescriptions basically amount to either excluding
occupation of Pauli forbidden two-body states in the three-body wave
function or excluding the apriori appearance of such undesired states.
The two prescriptions give somewhat different results and it is not yet
clear which one is preferable.  The projection technique, although
seems to be more straightforward, is in practice technically more
difficult.

The problem of treating the intercluster Pauli principle again
attracted attention in investigations of the structure of the
three-body halo nuclei \cite{zhu93} where most of the information 
is obtained by studying momentum distributions of fragments
of these nuclei in fast collisions with light targets.  Originally
these momentum distributions were expected to reveal the structure of
the initial three-body wave function \cite{kob88}. This is still true
although more elaborate analyses are required. The correct account for
the two-body interaction between fragments in the final state turned
out to be essential when low lying resonances are present
\cite{gar96,kor94}.

The purpose of the present paper is twofold. First we describe how the
projecting out technique can be easily incorporated in the adiabatic
hyperspherical expansion of the Faddeev equations in coordinate
space~\cite{fed93b}.  This method to solve the Faddeev equations has
already been used in several investigations of halo
nuclei~\cite{fed94a,fed94c,fed95}.  

Secondly we investigate how the differences in the prescriptions are
reflected in the momentum distributions of fragments from the break-up
reactions of three-body halos. This question is important for an
unambiguous interpretation of the wealth of experimental data on
three-body halos like $^6$He ($^4$He+n+n) and $^{11}$Li ($^9$Li+n+n).

After the introduction we shall in section 2 describe the method and
compare potentials and wave functions for two different
prescriptions. In section 3 we use these wave functions in
computations of momentum distributions in nuclear break-up reactions
of three-body halo nuclei. Finally in section 4 we give a brief summary
and the conclusions.

\section{Three-cluster problem with Pauli forbidden two-cluster
states} 


\paragraph*{Adiabatic hyperspherical expansion of the Faddeev equations.}
We shall use the hyperspherical coordinates ($\rho$, $\alpha$,
$\Omega_x$, $\Omega_y$) defined in refs. \cite{zhu93,fed94a}. The
volume element is given by $\rho^5 d\Omega d\rho$, where
$d\Omega=\sin^2\alpha \cos^2\alpha d\alpha d\Omega_x d\Omega_y$.  The
total wave function $\Psi_{J M}$ of the three-body system (with total
spin $J$ and projection $M$) is written as a sum of three components,
which in turn for each hyperradius $\rho$ are expanded in a complete
set of generalized angular functions $\Phi^{(i)}_{n}(\rho,\Omega_i)$
\begin{equation}
\Psi_{J M}= \frac {1}{\rho^{5/2}} 
  \sum_n f_n(\rho)  
\sum_{i=1}^3 \Phi^{(i)}_{n}(\rho ,\Omega_i) \; ,
\label{tot}
\end{equation}
where $\rho^{-5/2}$ is related to the volume element. 

These wave functions satisfy the angular part of the three Faddeev
equations:
\begin{equation} 
 {\hbar^2 \over 2m}\frac{1}{\rho^2}\hat\Lambda^2 \Phi^{(i)}_{n} 
 +V_{jk} (\Phi^{(i)}_{n}+\Phi^{(j)}_{n J M}+
                \Phi^{(k)}_{n})\equiv {\hbar^2 \over
2m}\frac{1}{\rho^2} \lambda_n(\rho) \Phi^{(i)}_{n}  \; ,
\label{ang}
\end{equation}
where $\{i,j,k\}$ is a cyclic permutation of $\{1,2,3\}$, $m$ is
an arbitrary normalization mass, and $\hat\Lambda^2$ is the
$\rho$-independent part of the kinetic energy operator. The analytic
expressions for $\hat{\Lambda}^2$ and the kinetic energy operator can
for instance be found in \cite{fed94a}.

The radial expansion coefficients $f_n(\rho)$ are obtained from 
a coupled set of ``radial'' differential equations \cite{fed94a}, i.e.
\begin{eqnarray} \label{rad} 
   \left(-\frac{\rm d ^2}{\rm d \rho^2}
   -{2mE\over\hbar^2}+ \frac{1}{\rho^2}\left( \lambda_n(\rho) - Q_{n n} +
  \frac{15}{4}\right) \right)f_n(\rho) =
 \nonumber \\ 
  \sum_{n' \neq n}   \left(
   2P_{n n'}{\rm d \over\rm d \rho}
   +Q_{n n'}
   \right)f_{n'}(\rho)  \; ,
\end{eqnarray}
where the functions $P$ and $Q$ are defined as angular integrals:
\begin{equation}
   P_{n n'}(\rho)\equiv \sum_{i,j=1}^{3}
   \int d\Omega \Phi_n^{(i)\ast}(\rho,\Omega)
   {\partial\over\partial\rho}\Phi_{n'}^{(j)}(\rho,\Omega)  \; ,
\end{equation}
\begin{equation}
   Q_{n n'}(\rho)\equiv \sum_{i,j=1}^{3}
   \int d\Omega \Phi_n^{(i)\ast}(\rho,\Omega)
   {\partial^2\over\partial\rho^2}\Phi_{n'}^{(j)}(\rho,\Omega)  \; .
\end{equation}

The radial equations in eq.(\ref{rad}) reveal that
$\hbar^2(\lambda_n(\rho) - Q_{n n} + \frac{15}{4})/(2m\rho^2)$ is the
diagonal part of the effective radial potential. Its behavior is
decisive for the properties of the three-body system. The
$\lambda$-spectrum at both $\rho = 0$ and $\rho = \infty$ is identical
to the hyperspherical spectrum. In addition, for every bound two-body
state, there exists one $\lambda$-value which bends over and diverges
parabolically as function of $\rho$ for $\rho \rightarrow \infty$, see
\cite{fed94c}. Such a level corresponds at large distances to the
three-body structure where the two-body subsystem is in the
corresponding bound state, whereas the third particle is far away.

\paragraph*{Treatment of the Pauli principle.}
The first prescription is a two-body potential without Pauli forbidden states
but with the same low energy properties as the original deep two-body
potential with the forbidden state. This corresponds to the use of an
additional Pauli repulsion.  We have discarded the apparently rigorous
procedure of constructing the phase equivalent potential from the
original two-body potential. The reasons are that the original
interaction in any case appears in the form of an effective (mean)
potential adjusted to reproduce specific measured quantities.
Furthermore, the binding energy of the three-body system must be
accurately reproduced to provide the correct structure. An additional
fine tuning is therefore almost inevitable. The strictly phase
equivalent potential is substantially more difficult to obtain and
use.  Under these circumstances we prefer to take the pragmatic
approach and construct phenomenological potentials adjusted to have
specifically chosen crucial properties.

The second prescription is a unique feature of the adiabatical
expansion method.  Since in this method each of the two-body bound
states gives rise to a separate diverging $\lambda$-value, the
prescription is simply to omit the $\lambda$-values corresponding to
the Pauli forbidden two-body bound states from the expansion
(\ref{tot}).  Then all the Pauli forbidden states are automatically
excluded in the three-body wave functions.

\section{Two-body potentials} 

We shall consider the two halo nuclei $^6$He and $^{11}$Li, both
approximated as three-body systems consisting of a core ($^4$He and
$^9$Li) surrounded by two valence neutrons.  This approximation works
remarkably well for these systems and furthermore due to relatively weak
binding of the systems only the low energy parameters of the two-body
potentials are essential for the structure of the systems~\cite{zhu93}.
As the core is inert in our approximation it is sufficient for our
purpose to assume zero spin of the core for $^{11}$Li.

In both cases the lowest neutron s-states in the core are filled.
Therefore to conform with the Pauli principle the halo neutron should
be prevented from occupying the lowest neutron-core s-state.  We
achieve this goal by use of two different neutron-core potentials with
the same low energy properties. One of the potentials will have a Pauli
forbidden bound state the other will not. This Pauli forbidden state
will be consequently removed from the active space.

\paragraph*{Neutron-neutron potential.} 

As indicated in \cite{zhu93} and proved by our test runs the particular
radial shape of the $n-n$ interaction is not important for $^6$He and
$^{11}$Li ground state properties as long as the low energy $n-n$
scattering parameters are correct. We therefore use a simple potential
similar to the one in \cite{fed95} which reproduces the experimental s-
and p-wave scattering lengths and effective ranges. It contains central,
spin-orbit (${\bf L \cdot S}$), tensor ($S_{12}$) and spin-spin (${\bf
s}_1\cdot{\bf s}_2$) interactions and is explicitly given as
\begin{eqnarray}V_{nn}(r)=37.05 \exp(-(r/1.31)^2)
	-7.38\exp(-(r/1.84)^2) \nonumber \\
	-23.77\exp(-(r/1.45)^2){\bf L \cdot S}
	+7.16\exp(-(r/2.43)^2) S_{12}\nonumber \\+
 \left(49.40\exp(-(r/1.31)^2)+
  29.53\exp(-(r/1.84)^2) \right) {\bf s}_1 \cdot {\bf s}_2,
\end{eqnarray}
where the strengths are in MeV and ranges in fm.  Its scattering
lengths\footnote{our sign convention is 
$k^{2l+1}\cot(\delta)\stackrel{k\rightarrow 0}{\rightarrow}
1/a+r_ek^2/2$} $a$ and effective ranges $r_e$ are (in fm) $a(^1S_0)$= 18.45,
$r_e(^1S_0)$= 2.83, $a(^3P_0)$= 3.38, $r_e(^3P_0)$= 1.10,
$a(^3P_1)$= $-2.02$, $r_e(^3P_1)$= $-2.94$, $a(^3P_2)$=
0.31, $r_e(^3P_2)$= 18.73.

\paragraph*{Neutron-$^4$He potential.} 

For $^6$He the low energy properties of the neutron-core subsystem in
the s- and p-waves are rather well known and reflected in an s-wave
scattering length of $a_s=-3.07 \pm 0.02$~fm and two resonance energies
$E$ and widths $\Gamma$ of $E(p_{3/2})=0.77$~MeV,
$\Gamma(p_{3/2})=0.64$~MeV and $E(p_{1/2}) \approx 1.97$~MeV,
$\Gamma(p_{1/2}) \approx 5.22$~MeV, see~\cite{ajz88}. The
$p_{1/2}$-$p_{3/2}$ splitting demands a spin-orbit force or
equivalently (with the same range of the interactions) different
strengths in the $p_{1/2}$ and $p_{3/2}$ channels.  The negative s-wave
scattering length can be reproduced by a repulsive potential or by an
attractive potential with a bound state which in this case is the Pauli
forbidden state occupied by the core neutrons.

We introduce central l-dependent and spin-orbit components in the
potential with a gaussian shape $S$exp$(-r^2/b^2)$. The range was
chosen to be $b=2.33$~fm for all components (similar to \cite{zhu93})
except for the repulsive s-wave where the range $b$ was changed to
3.34~fm  in order to reproduce the same effective range as for
attractive potential.  The strength parameters are then defined by the
fit to the specified scattering length and positions of the
resonances.

The p-wave strengths are $S(p_{1/2})=-48.675$~MeV and
$S(p_{3/2})=-53.175$~MeV (or, equivalently, the central p-wave strength
of -51.675~Mev and the spin-orbit strength of -3.00~MeV) which provide
the p-wave resonances $E(p_{1/2})=1.94$~MeV, $\Gamma(p_{1/2})=4.0$~MeV
and $E(p_{3/2})=0.77$~MeV, $\Gamma(p_{3/2})=0.73$~MeV. These parameters
are the same for both repulsive and attractive s-wave potential.

The s-wave strength for the attractive potential (further referred to
as ``attractive") is $S(s_{1/2})=-39.2$~MeV ($r_e$=1.41~fm and
$a=-3.07$~fm, one Pauli forbidden bound state). Without additional fit
these s- and p-wave potentials provide the binding energy
B($^6$He)=1.0~MeV and r.m.s. radius R($^6$He)=2.45~fm, which is close
to experimental data B($^6$He)=$0.97\pm$~0.04~MeV and
R($^6$He)$=2.57\pm~0.10$~fm.

The repulsive s-wave potential which reproduces the experimental
scattering length and the same effective range as the attractive
potential slightly underbindes $^6$He (by approximately 200~keV). Such
underbinding for repulsive potentials is not unusual for $^6$He.  To
alleviate this problem people normally increase the range of all
potentials by a few per cent~\cite{zhu93}. However this procedure
shifts the positions of the p-resonances from their experimental
values.  As we shall see below these positions are extremely important
for the momentum distributions.  We therefore leave the p-waves
unchanged and instead reduce slightly the repulsion in the s-wave.

The resulting potential is further referred to as ``repulsive" and has
the s-wave range and strength $b=3.34$~fm, $S(s_{1/2})=9.70$~MeV,
($a=-2.58$~fm, $r_e=$0.67~fm) which leads to the binding energy
B($^6$He)=1.0~MeV and r.m.s. radius R($^6$He)=2.50~fm.

\paragraph*{Neutron-$^9$Li potential.} 

For $^{11}$Li the low energy neutron-core data are less known, although
evidence is accumulating for a low lying virtual s-state at $E(s)
\approx 0.15 \pm 0.15$ MeV and the lowest p-resonance at $E(p) \approx
0.6 \pm 0.2$ MeV, see \cite{boh93,you94,zin95,abr95}.  The Pauli
forbidden states are in this case both the lowest s-state and the
$p_{3/2}$-state. For simplicity the latter is in the calculations
placed at high positive energy and thereby removed from the active
space by a large inverse spin-orbit potential \cite{fed95}.  For the
s-waves we use a shallow potential without bound state and a
deep potential with a Pauli forbidden state.

The radial shapes of the neutron-core interactions are also assumed to
be gaussians, i.e. $S$exp$(-r^2/b^2)$ with $b=2.55$~fm \cite{joh90}
except for the deep s-wave potential with a bound state where we use
$b=1.49$~fm to maintain the same effective range as for the shallow
potential.

We adopt further the usual assumption that the neutron-core
interactions do not depend on the spin of the $^9$Li-core.  A more
realistic study of $^{11}$Li properties taking into
account the spin of $^{9}$Li has been made in \cite{gar96,fed95}.

Again the spin-orbit neutron-core potential effectively only gives
different strengths for the two different p-waves. With the choice of
range for the radial potentials there are only three strength
parameters left each related to a resonance, a virtual state or a
scattering length. The spin-orbit force is used to remove the Pauli
forbidden $p_{3/2}$-state from the active space.  The two remaining
strength parameters then determine the s- and p$_{1/2}$-state as well
as the binding energy of $^{11}$Li.  One of them must be used to fine
tune the binding energy B of the total three-body system
B($^{11}$Li)$=0.295\pm~0.035$~MeV.  With correct binding energy the
root mean square radius is then always reproduced within the
experimental uncertainty, which is R($^{11}$Li)$=3.1\pm~0.3$~fm.

The potential with a bound Pauli forbidden s-state (further referred to
as ``deep") has the parameters $b=1.49$~fm for the s-wave and
$b=2.55$~fm for p-waves, $S(s_{1/2})=-176.608$~MeV (one bound state,
$a$=8.738~fm, virtual level at~0.20~MeV), $S(p_{3/2})=9.55$~MeV
($E(p_{3/2}$) is high and uninteresting) and $S(p_{1/2})=-38.34$~MeV
($E(p_{1/2})=0.77$~MeV, $\Gamma(p_{1/2})=0.89$~MeV).  The binding
energy and root mean square radius is then computed to be
B($^{11}$Li)=0.30~MeV and R($^{11}$Li)=3.34~fm.

The shallow potential without bound states but with the same low energy
properties as the deep potential overbinds $^{11}$Li by some 180~keV.
In contrast to $^6$He the dominating component is now the s-wave in the
neutron-core subsystem.  Therefore to fine tune the binding energy we
modify now the p-wave potential namely the position of the $p_{1/2}$
resonance while keeping the other properties unmodified.

The fine tuned potentials without a bound state (further referred to as
``shallow") has the parameters $b=2.55$ fm, $S(s_{1/2})=-7.14$~MeV (no
bound state, $a$=8.738~fm, virtual level at 0.20~MeV), 
$S(p_{3/2})=9.55$ MeV ($E(p_{3/2})$ is high)
and $S(p_{1/2})=-35.45$~MeV ($E(p_{1/2})=1.7$~MeV,
$\Gamma(p_{1/2})=4.2$~MeV).  The binding energy and root mean square
radius is now computed to be B($^{11}$Li)=~0.30 MeV and
R($^{11}$Li)=~3.35~fm.

\begin{figure}[t]
\epsfxsize=12cm
\epsfysize=7cm
\epsfbox[550 200 1100 550]{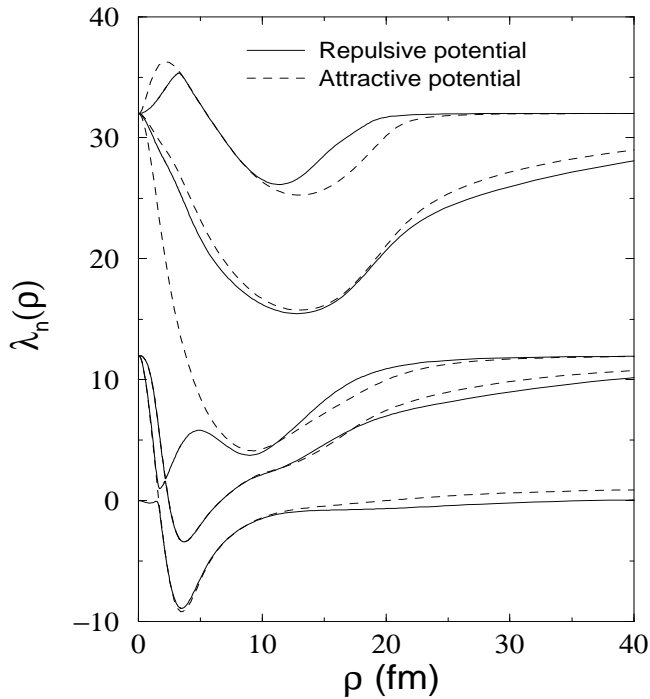}
\vspace{3.5cm}
\caption{\protect\small
Angular eigenvalue spectra $\lambda_n$ as
function of $\rho$ for $^6$He for the repulsive and attractive
neutron-$^4$He potentials described in the text.  The lowest
$\lambda$ for the attractive potential originating from zero and
corresponding to the Pauli forbidden state is removed.
}
\label{1a}
\end{figure}

\section{Effective three-body adiabatic potentials}

\paragraph*{$^6$He.} The effective radial potential in eq.(\ref{rad})
is the crucial quantity which in turn mainly is determined by the
$\lambda$-spectrum. The two prescriptions for dealing with the Pauli
principle lead to different spectra as seen in fig. \ref{1a} for $^6$He. The
purely repulsive neutron-core s-wave potential (solid curves) results
in the steep increase of the lowest $\lambda$-value at small $\rho$.
The increase is quickly interrupted by two (avoided) crossings
involving the next two $\lambda$-values, which in contrast are moving
steeply down due to the strong p-wave attraction leading to a low lying
p-resonance in $^5$He (see the parameters given in the previous section).
They therefore must contain a substantial amount of p-wave and the
related components must be dominating in the wave function which
accordingly also has about 87\% p-wave and 13\% s-wave in the relative
neutron-core system.  The curves eventually return back recovering the
hyperspherical spectrum for $\rho = \infty$. This occurs without
(avoided) level crossings.  The interaction between crossing levels is
vanishing or small indicating different symmetries of these levels. The
related almost preserved quantum numbers are the neutron-core relative
angular momenta $l_x=0$ and 1 corresponding to states originating from
zero and from 12.  The two levels originating from 32 are apparently at
small $\rho$ dominated respectively by s- and p-wave components. At
larger $\rho$ the highest $\lambda$ bends over due to p-wave admixture
and the related strong attraction.

The $\lambda$-spectrum for the other potential with sufficient
attraction in the neutron-core s-wave to support one bound s-state is
also shown as the dashed curves in fig. \ref{1a}. The lowest $\lambda$-value
is not shown in the figure. It is mainly s-waves in the neutron-core
subsystem and decreases at small $\rho$ and diverges parabolically for
$\rho \rightarrow \infty$ as clear signals of strong attraction and a
resulting bound state. All states built on this (not plotted) level
are Pauli forbidden and the level is therefore excluded in the
following calculations. The two next $\lambda$-values on the other
hand remain almost unchanged (apart from the avoided crossings)
indicating a dominating p-wave content (about 93\%) in the relative
neutron-core subsystem. This behavior is consistent with an almost
unchanged energy of the lowest lying p-resonance in the two-body
subsystem. As the lowest lying levels dominate the wave function the
behavior of these effective potentials also guarantee an almost
identical wavefunction in the two approaches.  Also the two highest
$\lambda$-values remain essentially unchanged. The third
$\lambda$-value start out as an s-wave as seen from the strong
decrease. At larger distances it is then replacing the s-wave
originating from zero (crossing two times) from the purely repulsive
potential.

\begin{figure}[t]
\epsfxsize=12cm
\epsfysize=7cm
\epsfbox[550 200 1100 550]{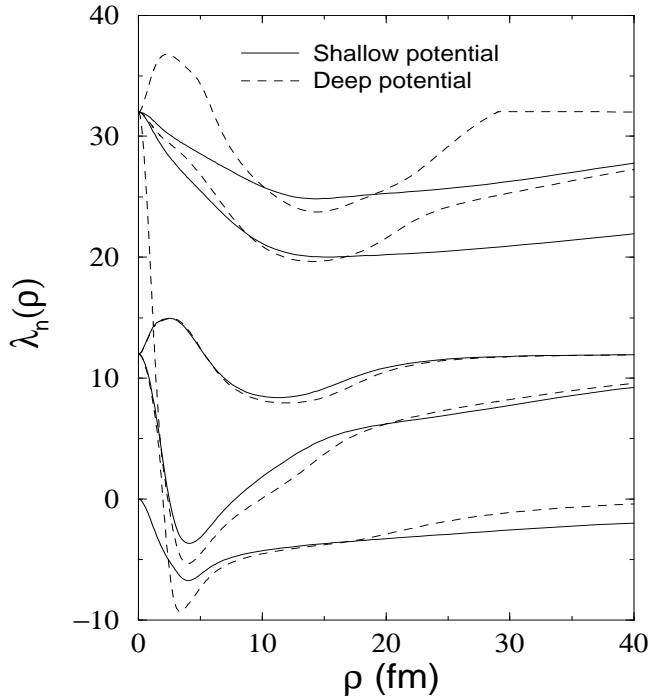}
\vspace{3.5cm}
\caption{\protect\small
Angular eigenvalue spectra $\lambda_n$ as
function of $\rho$ for $^{11}$Li for the shallow and deep neutron- $^{9}$Li
potentials described in the text. The lowest
$\lambda$ for the deep potential originating from zero and
corresponding to the Pauli forbidden state is removed.
}
\label{1b}
\end{figure}

\paragraph*{$^{11}$Li.}
For $^{11}$Li the low lying s-state of $^{10}$Li results in a
structure where the dominating neutron-core relative configuration is
an s-state.  This is rather different from $^6$He where the
predominant neutron-core relative configurations were p-states. The
$\lambda$-spectra for the two potentials corresponding to the
different prescriptions are shown in fig. \ref{1b}. The shallow attractive
potential  without a bound s-state (solid curves) results in a rather
slowly changing lowest $\lambda$-value which contains essentially only
neutron-core s-waves. This would by far carry the largest probability
in the resulting wave function. The two highest $\lambda$-values are
also smooth functions of $\rho$ while the second level steeply
decreases at small $\rho$ indicating attraction in the corresponding
partial wave.

The other potential, which has a Pauli forbidden bound neutron-core
s-state, has much stronger s-wave attraction and a $\lambda$-spectrum,
see dashed curves fig. \ref{1b}, with a fast moving level originating from
32. This level responds to the large s-wave attraction and crosses
quickly the two levels originating from 12 before replacing the lowest
level from the shallow potential. This level will dominate the
configuration of the radial wave function.  The lowest $\lambda$-value
originating from zero is not plotted and also omitted in the following
calculations. It decreases at small $\rho$ and diverges parabolically
at large $\rho$.  The four remaining levels all have similarly
behaving counter parts in the other spectrum.  Again we see the
(avoided) crossings indicating almost conserved quantum numbers on the
levels. As before they can be traced back to the neutron-core relative
angular momentum $l_x=0$ and 1.

\begin{figure}[t]
\epsfxsize=12cm
\epsfysize=7cm
\epsfbox[550 200 1100 550]{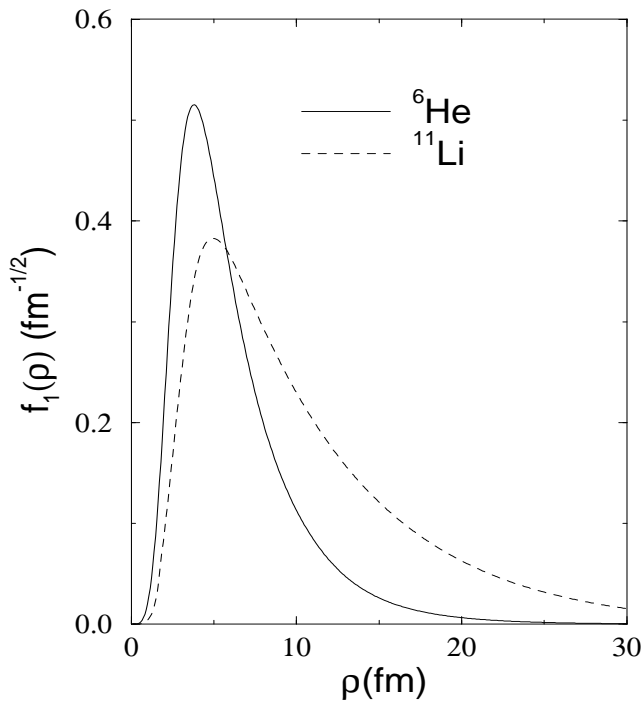}
\vspace{3.5cm}
\caption{\protect\small
The radial wave functions for $^6$He (solid
curve) and $^{11}$Li (dashed curve) corresponding to the lowest
dominating $\lambda$-values for the potentials with Pauli forbidden
bound states.  }
\label{2}
\end{figure}

\paragraph*{Radial functions.}
The radial wave function corresponding to the lowest $\lambda$ is shown
in fig. \ref{2} for the two nuclei. This component is responsible for more
than 90\% of the wave function.  Only the result of one of the
prescriptions is shown for each case, since the curves would be
difficult to distinguish especially at larger distances where they are
very close.  This is closely related to the fact that the binding
energies and radii are the same in both prescriptions.  The wave
function is spatially more extended for $^{11}$Li than for $^6$He,
again reflecting the difference in binding energy.

\section{Momentum distributions} 

The momentum distributions after fragmentation of halo nuclei is a
very direct way of gaining detailed information about the relative
wave function of these nuclei. It was originally expected simply to
provide the Fourier transform, but more complicated and detailed
analyses are needed. Such computations are fortunately available at
this moment and we shall exploit this avenue to test our
prescriptions.

\paragraph*{The sudden approximation.}

We consider a process where a high-energy three-body halo projectile
instantaneously looses one of the particles without disturbing the
remaining two. We also assume a light target and we shall therefore
neglect the Coulomb dissociation process, which then only contributes
marginally.

We work in the center of mass system of the three-body projectile and
denote by $\mbox{\bf k}_y$ and $\mbox{\bf k}_x$ the total and relative
momentum of the two remaining particles in the final state. The
transition matrix of the reaction in the sudden approximation is then
given by
\begin{equation}
M^{JM}_{s_x \sigma_x s_y \sigma_y}(\mbox{\bf k}_y, \mbox{\bf k}_x) 
 \propto \langle
 e^{i \mbox{\scriptsize \bf k}_y \cdot \mbox{\scriptsize \bf y} }
  \chi_{s_y \sigma_y} w_{s_x \sigma_x}(\mbox{\bf k}_x, \mbox{\bf x})
           | \Psi_{JM} \rangle  \; ,
\label{tran}
\end{equation}
where $\Psi_{JM}$ is the three-body wave function and $w_{s_x
\sigma_x}$ is the final state {\it distorted} two-body wave function
\cite{gar96} corresponding to the two remaining particles with the
distance {\bf x}, total spin and projection equal to $s_x$ and
$\sigma_x$. The distance between the center of mass of the two-body
system and the removed particle is {\bf y} and $\chi_{s_y \sigma_y}$ is
the spin wave function of the third particle where $s_y$ and $\sigma_y$
are the related total spin and projection.

The cross section or momentum distribution is now obtained by squaring
the transition matrix and subsequently averaging over initial states
and summing over final states:  
\begin{equation}
\frac{d^6\sigma}{d\mbox{\bf k}_x d\mbox{\bf k}_y} \propto
\sum_M \sum_{s_x \sigma_x \sigma_y}
|M^{J M}_{s_x \sigma_x s_y \sigma_y}(\mbox{\bf k}_x, \mbox{\bf k}_y)|^2 \; .
\label{mom}
\end{equation}

Using the momentum $\mbox{\bf p} (=a \mbox{\bf k}_x + b \mbox{\bf
k}_y)$ of one of the particles relative to the center of mass of the
projectile as the variable instead of $\mbox{\bf k}_x$, we obtain the
relation
\begin{equation}
\frac{d^6\sigma}{d\mbox{\bf p} d\mbox{\bf k}_y} = 
  \frac{1}{a^3} \frac{d^6\sigma}{d\mbox{\bf k}_x d\mbox{\bf k}_y}
  \propto \frac{1}{a^3}
\sum_M \sum_{s_x \sigma_x \sigma_y}
|M^{J M}_{s_x \sigma_x s_y \sigma_y}(\mbox{\bf k}_x,
 \mbox{\bf k}_y)|^2 \; ,
\label{rot}
\end{equation}
where $a^3$ arises from the Jacobi determinant for the transformation.
The differential cross section in eq.(\ref{rot}) should be integrated
over all unobserved variables, i.e. $\mbox{\bf k}_y$ and some of the
components of {\bf p}. Note that we have not specified any coordinate
system, and the axis {\bf x}, {\bf y}, and {\bf z} are therefore
completely arbitrary. Thus, in our approximation the longitudinal and
transverse momentum distributions are identical.

After neutron removal fragmentation reactions is usual to define the
invariant mass $E_{\mbox{\scriptsize core+n}}$ as
\begin{equation}
E_{\mbox{\scriptsize core+n}}=\left( (E_{\mbox{\scriptsize core}}+
E_{\mbox{\scriptsize n}})^2 + c^2 (\mbox{\bf p}_{\mbox{\scriptsize
core}}+\mbox{\bf p}_{\mbox{\scriptsize n}})^2 \right)^{1/2} -
(M_{\mbox{\scriptsize core}} + M_{\mbox{\scriptsize n}})c^2, 
\label{invdef}
\end{equation}
where $E_{\mbox{\scriptsize core,n}}$, $\mbox{\bf p}_{\mbox{\scriptsize
core,n}}$, and $M_{\mbox{\scriptsize core,n}}$ denote the energy,
three-momentum, and rest mass of the core and the neutron,
respectively. 

Computing the invariant mass in the frame of the two-body system after
the fragmentation ($\mbox{\bf p}_{\mbox{\scriptsize core}}+\mbox{\bf
p}_{\mbox{\scriptsize n}}=0$) we have
\begin{equation}
p_{\mbox{\scriptsize core}}^2 c^2 = p_{\mbox{\scriptsize n}}^2 c^2 =
\frac{1}{m} \frac{M_{\mbox{\scriptsize core}} M_{\mbox{\scriptsize n}}}
{M_{\mbox{\scriptsize core}} + M_{\mbox{\scriptsize n}}} k_x^2 c^2 =
E_{\mbox{\scriptsize core}}^2 - M_{\mbox{\scriptsize core}}^2 c^4 =
E_{\mbox{\scriptsize n}}^2 - M_{\mbox{\scriptsize n}}^2 c^4,
\end{equation}
and
\begin{equation}
dE_{\mbox{\scriptsize core+n}} = dE_{\mbox{\scriptsize core}} +
dE_{\mbox{\scriptsize n}} = \frac{E_{\mbox{\scriptsize core}} +
E_{\mbox{\scriptsize n}} }{E_{\mbox{\scriptsize core}} 
E_{\mbox{\scriptsize n}} } 
\frac{M_{\mbox{\scriptsize core}} M_{\mbox{\scriptsize n}}}
     {m (M_{\mbox{\scriptsize core}}+M_{\mbox{\scriptsize n}})}
     k_x dk_x
\end{equation}

The invariant mass spectrum is then defined
\begin{equation}
\frac{d\sigma}{dE_{\mbox{\scriptsize core+n}}}=
\frac{E_{\mbox{\scriptsize core}} E_{\mbox{\scriptsize n}} }
{E_{\mbox{\scriptsize core}} + E_{\mbox{\scriptsize n}} }
\frac{m (M_{\mbox{\scriptsize core}} + M_{\mbox{\scriptsize n}})}
{M_{\mbox{\scriptsize core}} M_{\mbox{\scriptsize n}}}
\frac{1}{k_x} \frac{d\sigma}{dk_x}
\label{invsp}
\end{equation}
where $d\sigma/dk_x$ is obtained from eq.(\ref{mom}) after integrating
over the unobserved quantities.

\begin{figure}[t]
\epsfxsize=12cm
\epsfysize=7cm
\epsfbox[550 200 1100 550]{prep4_3.ps}
\vspace{2.9cm}
\caption{\protect\small Longitudinal core momentum distributions for a
$^6$He neutron removal reaction computed for both the purely repulsive
potential (solid curve) and the attractive potential with one bound
s-state (dashed curve) defined in fig. \ref{1a}. The broadest distributions
correspond to the neglect of final state interaction. The experimental
points for fast $^6$He (400 MeV/u) colliding with a carbon target are
from \protect\cite{koba92}, and correspond to the transverse core momentum
distribution. The core momentum is referred to the center of mass
system of the three-body projectile.  }
\label{3}
\end{figure}
\vspace{-12mm}
\begin{figure}[t]
\epsfxsize=12cm
\epsfysize=7cm
\epsfbox[550 200 1100 550]{prep4_4a.ps}
\vspace{2.9cm}
\caption{\protect\small
Longitudinal neutron momentum distributions for
a $^6$He neutron removal reaction computed for both the repulsive
potential (solid curve) and the attractive potential with one bound
s-state (dashed curve) defined in the text.  The broadest distributions
correspond to the neglect of final state interaction. The experimental
points for fast $^6$He (800 MeV/u) colliding with a carbon target are
from \protect\cite{koba93}, and correspond to the transverse neutron momentum
distribution. The neutron momentum is referred to the center of
mass system of the three-body projectile.
}
\label{4a}
\end{figure}

\begin{figure}[t]
\epsfxsize=12cm
\epsfysize=7cm
\epsfbox[550 200 1100 550]{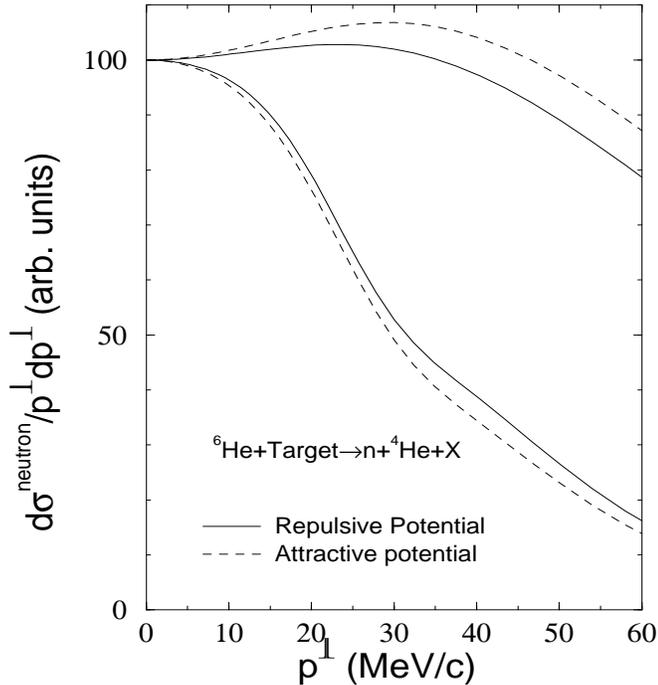}
\vspace{3.5cm}
\caption{\protect\small
Radial neutron momentum distributions for a
$^6$He neutron removal reaction computed for both the repulsive
potential (solid curve) and the attractive potential with one bound
s-state (dashed curve) defined in the text.  The broadest distributions
correspond to the neglect of final state interaction.  The neutron
momentum is referred to the center of mass system of the three-body
projectile.
}
\label{4b}
\end{figure}

\paragraph*{$^6$He fragmentation.} 

In fig. \ref{3} the computed longitudinal core momentum distribution after
neutron removal in a fast nuclear break-up reaction of $^6$He on a
light target is shown. The results of the two prescriptions are
exhibited both with and without inclusion of the final state
interaction.  As expected the effect of the final state interaction is
visible but not extremely important even with this fairly light core.
The results of the two prescriptions are very close when the final
state interaction is included. When it is neglected the intrinsic
differences in the wave functions obtained with the two different
prescriptions to account for the Pauli principle produce a clear
distinction. The difference is in the additional node of the forbidden
state prescription compared to the reduced attraction prescription.
The experimental data in fig. \ref{3} are taken from~\cite{koba92}, and
correspond to the transverse core momentum distribution after fast
$^6$He fragmentation on a carbon target.  The present model
computations do not distinguish between directions, but the results are
expected to be more appropriate for the longitudinal directions, which
unfortunately are unavailable at this moment. However, the agreement
with the measured results is still encouraging for several reasons.
First, we have basically no free parameters. Second, the experimental
transverse momentum distributions are expected to be broader than the
longitudinal ones \cite{hum95}. Third, no broadening due to neglected
effects and experimental resolution is included in the computation.

In figs. \ref{4a} and \ref{4b} we show corresponding computed longitudinal and
radial ($p^\bot = (p_x^2 + p_y^2)^{1/2}$) neutron momentum
distributions for a neutron removal $^6$He fragmentation reaction.
Again the two prescriptions give very similar results when the final
state interaction is included. In this case we obtain a very large
influence of the final state interaction which reduce the full width at
half maximum by a factor 2-3. Inclusion of the final state interaction
is therefore necessary to obtain the observed agreement with the
measured values. We want here to emphasize that the computed results
are found in a consistent model where the same two-body potential is
responsible for both the three-body structure of the initial halo state
and the final state interaction after break-up.  The experimental data
in fig.~\ref{4a} are obtained from \cite{koba93}, and correspond to the
transverse neutron momentum distribution after $^6$He fragmentation on
a carbon target at 800 MeV/u. Again the computed curve is expected to
compare more favourably with the experimental longitudinal neutron
momentum distribution. However, these data are not presently
available.

\begin{figure}[t]
\epsfxsize=12cm
\epsfysize=7cm
\epsfbox[550 200 1100 550]{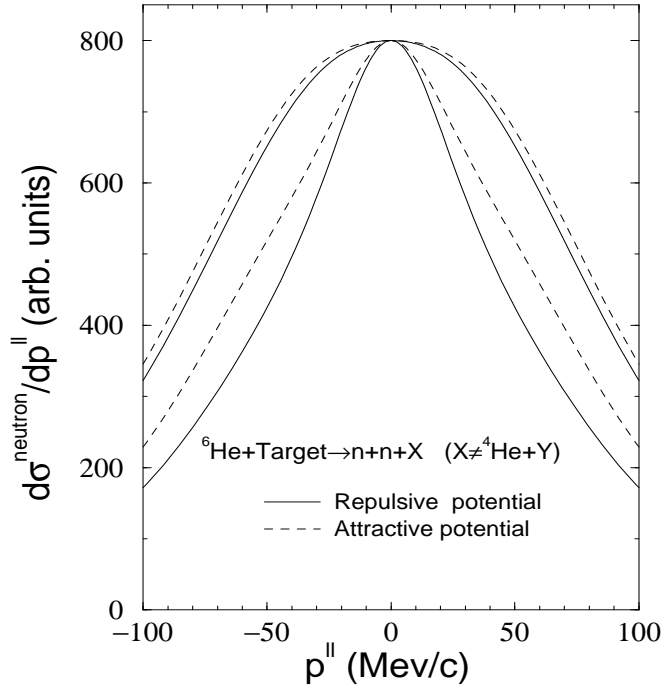}
\vspace{3.1cm}
\caption{\protect\small
Longitudinal neutron momentum distributions
for a $^6$He core break-up reaction computed for both the repulsive
potential (solid curve) and the attractive potential with one bound
s-state (dashed curve) defined in the text.  The broadest distributions
correspond to the neglect of final state interaction.  The neutron
momentum is referred to the center of mass system of the three-body
projectile.
}
\label{5a}
\end{figure}
\vspace{-10mm}
\begin{figure}[t]
\epsfxsize=12cm
\epsfysize=7cm
\epsfbox[550 200 1100 550]{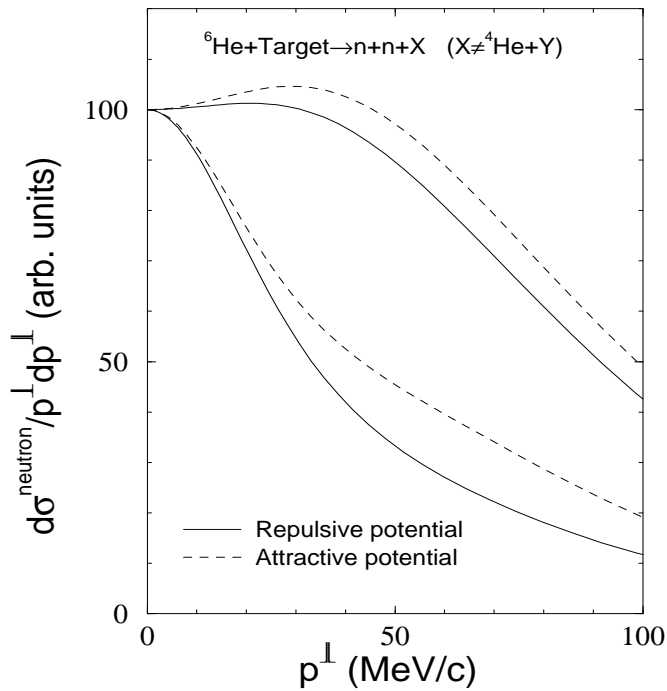}
\vspace{3.1cm}
\caption{\protect\small
Radial neutron momentum distributions
for a $^6$He core break-up reaction computed for both the repulsive
potential (solid curve) and the attractive potential with one bound
s-state (dashed curve) defined in the text.  The broadest distributions
correspond to the neglect of final state interaction.  The neutron
momentum is referred to the center of mass system of the three-body
projectile.
}
\label{5b}
\end{figure}

Although measured values are not available we show in figs. \ref{5a} and 
\ref{5b}
the computed longitudinal and radial neutron momentum distributions for
core break-up reactions of $^6$He. In this case the core is violently
removed, and the final neutron-neutron interaction is relevant. The two
prescriptions again lead to similar results. The final state
interaction can not be neglected due to the low lying virtual s-state
in the neutron-neutron system.

Key quantities for a number of different cases are given in table I,
where the $^6$He binding energy and radius are kept at the measured
values. For the case of the repulsive s-wave potential we have first
considered two cases (first two rows of table I) where both the
$p_{1/2}$ and the $p_{3/2}$-resonance energies lie within the
experimental values. In these two cases the computed scattering length
for the s-wave is smaller than the experimental value of $-3.07$.
Increasing the absolute size of the computed s-wave scattering length
(keeping fixed the energy of the lowest $p_{3/2}$-resonance) reduces
the energy of the $p_{1/2}$-resonance (case 3 in the table). When the
scattering length is equal to the experimental value the energy of the
second p-resonance is below the experimental energy. Cases 4 to
6 in table I are analogous to cases 1 to 3, but with an s-wave
repulsive potential giving a clearly worse agreement with the
experimental s-wave scattering length. However the changes in the full
width at half maximum of the momentum distributions, the p-wave content
and root mean square radius, are very small, reflecting that the
properties the s-wave interaction do not play an essential role in the
structure of $^6$He.

For an attractive s-potential with one bound state (two last rows of
table I) it is possible to reproduce both the experimental s-wave
scattering length and the two p-wave resonance energies. For the two
cases 7 and 8 shown in table I the variation of the energy of the
$p_{1/2}$-resonance only produces a change in the $^6$He binding energy
of about 70 keV, but in both cases consistent with a computed value of
1.0 MeV. The cases 1 and 7 correspond to the parameters given in
section 3, and they are the two cases considered for $^6$He in figs. \ref{3},
\ref{4a}, \ref{4b}, \ref{5a}, and \ref{5b}.

\paragraph*{$^{11}$Li fragmentation.} 

The interaction parameters for the schematic $^{11}$Li computation,
where the core spin is assumed to be zero, only provide the s- and
$p_{1/2}$-strengths for fine tuning. We already placed the
$p_{3/2}$-state at a very high energy and thereby removed it from the
active space. In contrast to $^6$He the dominating component is now an
s-wave in the neutron-core subsystem.   In table II we show the
relevant quantities for three different computations. The first row
corresponds to a computation with a shallow s-wave potential without
bound states while the second and third rows correspond to 
computations with deep s-wave potentials with one bound state. The
effective range and scattering length are the same in the three cases
and the s-wave interaction places the lowest virtual s-state at 200
keV.  Once the s-wave potential has been fixed the energy of the
$p_{1/2}$-resonance must then be around 1.7 MeV for the shallow
potential case in order to fit the $^{11}$Li binding energy.  When the
deep potential is used we consider two situations. First we keep the
energy of the $p_{3/2}$-resonance unchanged. Then the $p_{1/2}$-energy
is reduced to 0.8 MeV in order to recover the $^{11}$Li binding energy.
In the second case we force the $p_{1/2}$-resonance energy to take the
same value as for the shallow potential. The correct $^{11}$Li binding
energy is in this case obtained by reducing the energy of the
$p_{3/2}$-resonance. The cases 1 and 2 correspond to the parameters
given in section 3 and they will be the two cases considered for
$^{11}$Li in the next figures (more elaborated and detailed
computations for $^{11}$Li fragmentation reactions may be found
elsewhere, see \cite{gar96}).

\begin{table}
\begin{small}
\begin{tabular}{c|cc|cc|cc|cc}
case  &  $E(s_{1/2})$  &  $a$  & $E(p_{1/2})$ & $\Gamma(p_{1/2})$ &
  $\Gamma_c$ & $\Gamma_n$ & 
         $p$-content & R \\
  &  (MeV)  &  (fm)  &  (MeV)  &  (MeV)  &    
            (MeV/c)  & (MeV/c)  &  (\%)  & (fm) \\
                            \hline
 1  &  5.27  &  $-2.58$  &  1.94  &  4.02   &  101.6
& 72.6  &  87.7  & 2.50\\
 2  & 5.13 & $-2.21$ & 2.55  & 7.38  & 100.5 &  74.8 &
86.2  &  2.50\\
 3 &  5.09 & $-3.07$ & 1.28 & 1.77  & 103.0 & 68.1 & 89.4 & 2.50\\
 4  &  10.6  &  $-2.04$  &  1.94  &  4.02   &  98.8
& 72.9  &  86.7  & 2.53\\
 5  & 10.8 & $-1.80$ & 2.55  & 7.38  & 98.3 &  74.5 &
85.5  &  2.53\\
 6 &  9.3 & $-2.51$ & 1.10 & 1.34  & 99.2 & 67.1 & 88.5 & 2.53\\
                          \hline
 7 &  5.08  &  $-3.07$  & 1.94  &  4.02   & 107.6 & 67.4
& 92.8 & 2.45 \\
 8 &   5.08  &  $-3.07$  & 2.55  &  7.38  & 106.2 & 68.1
& 92.7 & 2.49 
\end{tabular}
\end{small}
\vspace{0.3cm}
\caption{ Key quantities for $^{6}$He corresponding to various
neutron-core interactions which are chosen as gaussians in each partial
wave. The neutron-neutron interaction is given in Section 3.  For the
neutron-$^4$He subsystem we give energies of the virtual
$s_{1/2}$-state, s-wave scattering lengths, energies and widths of the
$p_{1/2}$-resonance.  The $p_{3/2}$-resonance energy and width are in
all cases 0.77 MeV and 0.73 MeV, respectively. The full width at half
maximum for core and neutron momentum distributions is denoted
$\Gamma_c$ and $\Gamma _n$. We also give the probability for finding
the neutron-$^{4}$He subsystem in a $p$-wave in the three-body wave
function of $^{6}$He. The remaining probablity is found in s-waves. The
total binding energy of the three-body system is in all cases equal to
about 1.0 MeV and the root mean square radius is given in the last
column.  Cases 1-6 and 7-8 correspond to potentials without bound
and with one bound s-state, respectively.} 
\end{table}

Looking at the root mean square radius we observe that it only varies
between about 3.30 fm and 3.40 fm independently of the prescription and
mainly sensitive to the amount of s-state in the neutron-core
subsystem, see also \cite{fed95}.  The deep potential gives almost the
same radius as the shallow potential. This is in contradiction with the
large radii ($\geq$ 3.50 fm) obtained in similar Faddeev calculations,
where the lowest s-state is projected out \cite{tho94}. These
surprisingly large radii are hard to reconcile with the general
asymptotic relation between binding energy and radius for weakly bound
halo nuclei \cite{fed94a,fed94c}.  Incidentally a similarly large
radius ($3.55 \pm 0.10$ fm) is also obtained in recent Glauber analyses
of the measured reaction cross section \cite{alk96}. The reliability of
such analyses still remains to be investigated.

\begin{figure}[t]
\epsfxsize=12cm
\epsfysize=7cm
\epsfbox[550 200 1100 550]{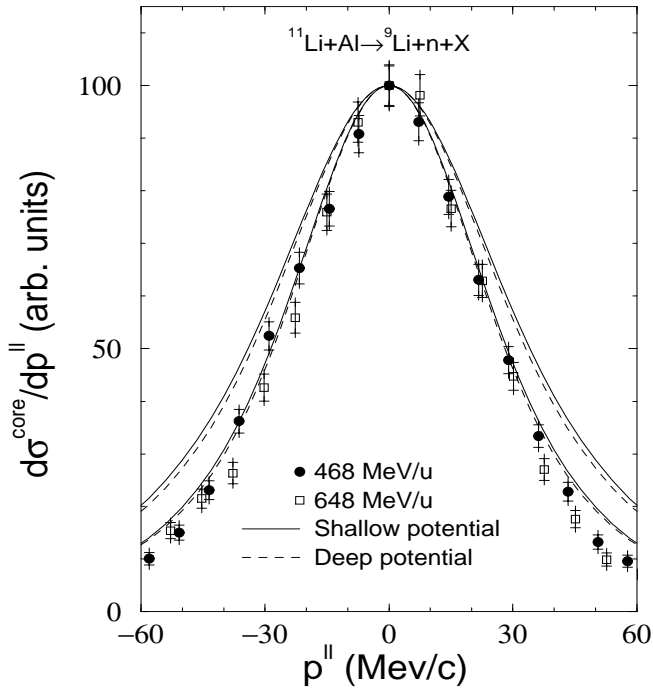}
\vspace{3.5cm}
\caption{\protect\small
Longitudinal core momentum distributions for a
$^{11}$Li neutron removal reaction computed for both the shallow
potential (solid curve) and the deep potential with one bound s-state
(dashed curve) defined in the text.  The broadest distributions
correspond to the neglect of final state interaction. The experimental
points for fast $^{11}$Li (468 MeV/u and 648 MeV/u) colliding with a
aluminium target are from \protect\cite{geiss96}.  The core momentum is
referred to the center of mass system of the three-body projectile.
}
\label{6}
\end{figure}

\begin{table}
\begin{tabular}{c|cc|cc|cc|cc}
case  &  $E(s_{1/2})$  &  $a$  & $E(p_{1/2})$ & $\Gamma(p_{1/2})$ &
     $\Gamma_c$ & $\Gamma_n$ & $p$-content & R  \\
  &  (MeV)  &  (fm)  &  (MeV)  &  (MeV)  &  (MeV/c)  & (MeV/c)  &  (\%) 
  &  (fm) \\
                            \hline
 1 &  0.200  &  8.738  &  1.68  &  4.16  &  57.6  &  34.8  &  17.3 & 3.34\\
                          \hline
 2 &  0.200  &  8.738  &  0.77  &  0.89  &  56.1  &  38.2  &  21.0 
                                                                 &  3.35\\
 3 &  0.200  &  8.738  &  1.68  &  4.16  &  55.6  &  35.6  &  15.6
                                                                 &  3.37
\end{tabular}
\vspace{0.3cm}
\caption{ Key quantities for $^{11}$Li corresponding to various
neutron-core interactions which are chosen as gaussians in each partial
wave. The neutron-neutron interaction is given in section 3.  For the
neutron-$^{9}$Li subsystem we give energies of the virtual
$s_{1/2}$-state, s-wave scattering lengths, energies and widths of the
$p_{1/2}$-resonance. The width and energy of the $p_{3/2}$-resonance is
unphysical and not included.  The full width at half maximum for core
and neutron momentum distributions is denoted $\Gamma_c$ and $\Gamma
_n$. We also give the probability for finding the neutron-$^{10}$Li
subsystem in a $p$-wave in the three-body wave function of $^{11}$Li.
The remaining probability is found in s-waves. The total binding energy
of the three-body system is in all cases equal to about 0.30 MeV and
the root mean square radius is given in the last column.  Cases 2 and 3
correspond to potentials with one bound s-state. } 
\end{table}

\begin{figure}[t]
\epsfxsize=12cm
\epsfysize=7cm
\epsfbox[550 200 1100 550]{prep4_7a.ps}
\vspace{3.1cm}
\caption{\protect\small
Longitudinal neutron momentum distributions for
a $^{11}$Li neutron removal reaction computed for both the shallow
potential (solid curve) and the deep potential with one bound s-state
(dashed curve) defined in the text. The broadest distributions
correspond to the neglect of final state interaction.  The neutron
momentum is referred to the center of mass system of the three-body
projectile.
}
\label{7a}
\end{figure}
\vspace{-10mm}
\begin{figure}[t]
\epsfxsize=12cm
\epsfysize=7cm
\epsfbox[550 200 1100 550]{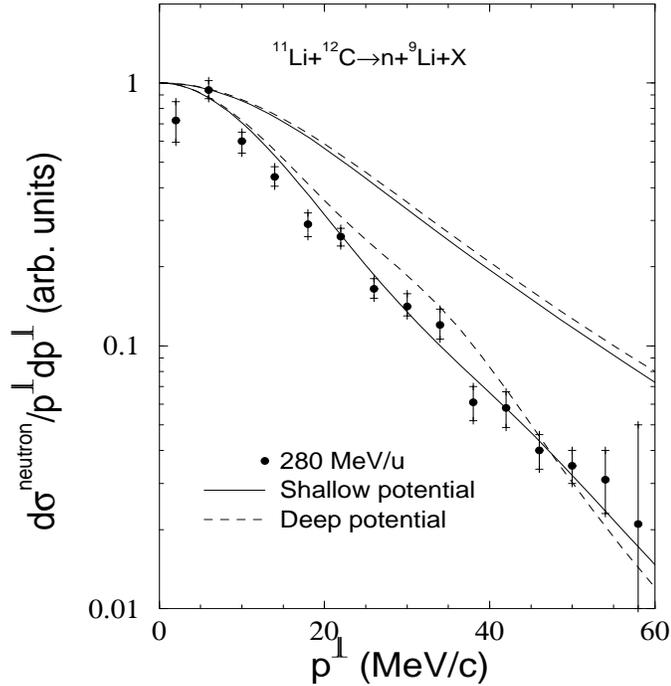}
\vspace{3.1cm}
\caption{\protect\small
Radial neutron momentum distributions for
a $^{11}$Li neutron removal reaction computed for both the shallow
potential (solid curve) and the deep potential with one bound s-state
(dashed curve) defined in the text. The broadest distributions
correspond to the neglect of final state interaction. The experimental
points for fast $^{11}$Li (280 MeV/u) colliding with a carbon target
are from \protect\cite{zin95}.  The neutron momentum is referred to the center
of mass system of the three-body projectile.
}
\label{7b}
\end{figure}

In fig. \ref{6} we show the computed longitudinal core momentum distributions
after neutron removal in a fast nuclear break-up reaction of $^{11}$Li
on a light target. We assume that the $^{9}$Li-core has spin
zero. The model is then without the proper spin couplings and the spin
splitting of the neutron-core relative states \cite{fed95}. Although
simplified the model is still not far from being realistic. The two
prescriptions again give very similar results and the effect of the
final state interaction is now a little smaller than for $^{6}$He due
to the larger mass of the core nucleus. The experimental distributions are
essentially reproduced, and correspond to a $^{11}$Li fragmentation
reaction on Al at 468 MeV/u and 648 MeV/u \cite{geiss96}.

In figs. \ref{7a} and \ref{7b} we show computed longitudinal and 
radial ($p^\bot
=(p^2_x+p^2_y)^{1/2}$) neutron momentum distributions for a neutron
removal reaction of $^{11}$Li. The cases shown in the figure correspond
to rows 1 and 2 of table II.  The difference in energy of the $p_{1/2}$
resonance in the neutron-core subsystem produces the difference in the
momentum distribution when the final state interaction is included.  In
fact the momentum distributions obtained in the third case of table II
are almost indistinguishable from those obtained in the first case (the
energy of the $p_{1/2}$ resonance is the same in these two cases).  The
final state interaction is then important due to the presence of low
lying virtual s-states and p-resonances, and very sensitive to the
energies of these low lying states.

\begin{figure}[t]
\epsfxsize=12cm
\epsfysize=7cm
\epsfbox[550 200 1100 550]{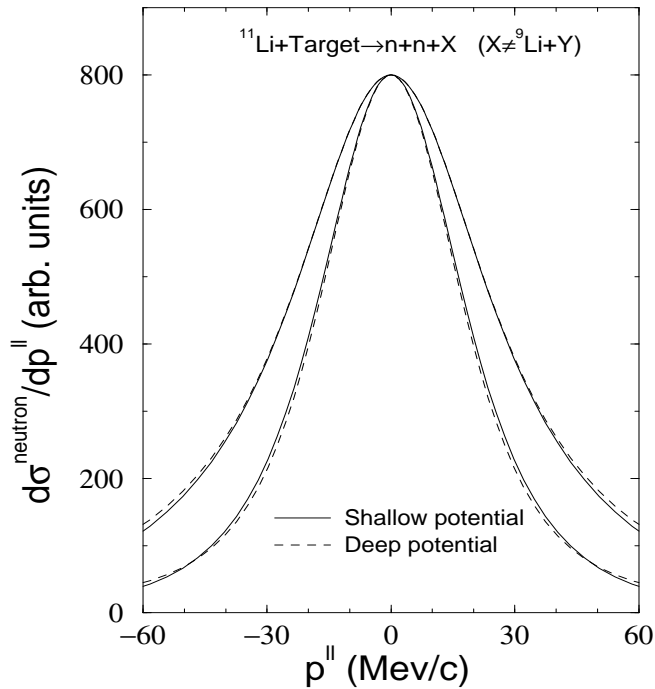}
\vspace{3.5cm}
\caption{\protect\small
Longitudinal neutron momentum distributions for
a $^{11}$Li core break-up reaction computed for both the shallow
potential (solid curve) and the deep potential with one bound
s-state (dashed curve) defined in the text.  The broadest distributions
correspond to the neglect of final state interaction. The neutron
momentum is referred to the center of mass system of the three-body
projectile.
}
\label{8a}
\end{figure}

\begin{figure}[t]
\epsfxsize=12cm
\epsfysize=7cm
\epsfbox[550 200 1100 550]{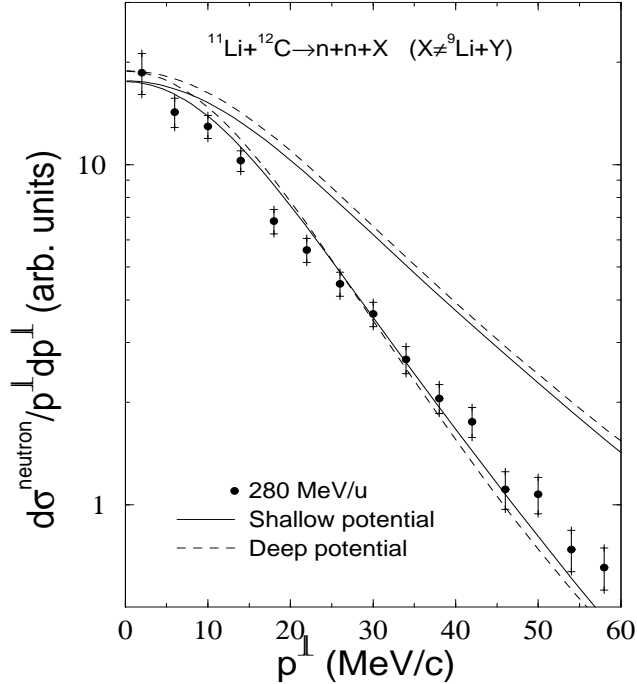}
\vspace{3.5cm}
\caption{\protect\small
Radial neutron momentum distributions for
a $^{11}$Li core break-up reaction computed for both the shallow
potential (solid curve) and the deep potential with one bound s-state
(dashed curve) defined in the text.  The broadest distributions
correspond to the neglect of final state interaction. The experimental
points for fast $^{11}$Li (280 MeV/u) colliding with a carbon target
are from \protect\cite{nil95}. The neutron momentum is referred to the center
of mass system of the three-body projectile.
}
\label{8b}
\end{figure}

For completeness we show in figs. \ref{8a} and \ref{8b} the longitudinal 
and radial
neutron momentum distribution from a $^{11}$Li core break-up reaction.
We again observe an important effect produced by the inclusion of the
final neutron-neutron interaction (the neutron-neutron interaction has a
low lying virtual s-state), and an almost identical result for
both the shallow and deep neutron-$^9$Li s-wave potential. In figs. \ref{7b}
and \ref{8b} the experimental data are fairly well reproduced, and correspond
to a $^{11}$Li fragmentation reaction at 280 MeV/u on a carbon target
\cite{zin95,nil95}

\paragraph*{Invariant mass spectrum.} 

Computing the invariant mass $E_{\mbox{\scriptsize core+n}}$
(eq.(\ref{invdef})) in the frame of the two-body system in the final
state ($\mbox{\bf p}_{\mbox{\scriptsize core}}+\mbox{\bf
p}_{\mbox{\scriptsize n}}=0$) we can interpret $E_{\mbox{\scriptsize
core+n}}$ as the kinetic energy of the neutron-core system in the final
state:
\begin{equation}
E_{\mbox{\scriptsize core+n}} \approx \frac{p_{\mbox{\scriptsize
core}}^2}{2 \mu} = \frac{p_{\mbox{\scriptsize n}}^2}{2 \mu}
= \frac{k_x^2}{2 m}
\end{equation}
where $\mu$ and $m$ are the reduced mass of the two-body system and the
arbitrary normalization mass, respectively. 
The energy of a resonance is defined as the $k_x^2/2 m$
value for which the cross section has a maximum.
As a consequence the invariant mass spectrum
(\ref{invsp}) will show a peak at the $E_{\mbox{\scriptsize core+n}}$ 
energy equal to a resonance energy.

\begin{figure}[t]
\epsfxsize=12cm
\epsfysize=7cm
\epsfbox[550 200 1100 550]{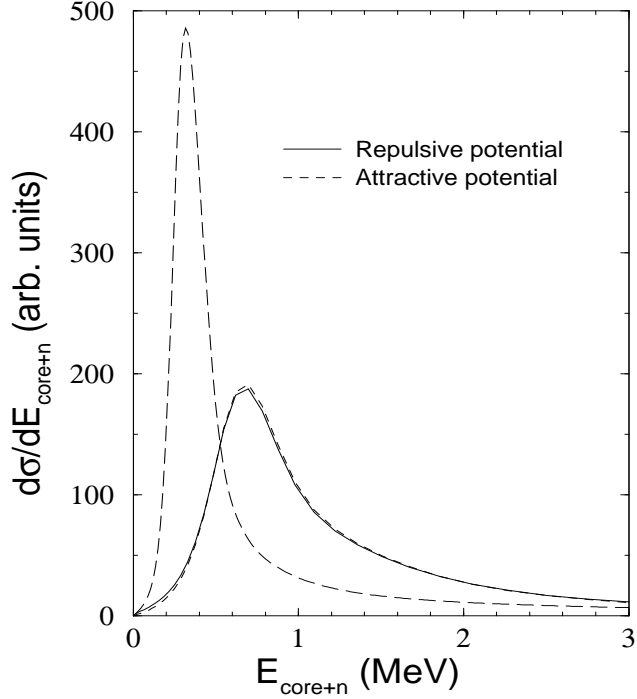}
\vspace{3.1cm}
\caption{\protect\small
The invariant mass spectrum for a $^6$He neutron
removal  reaction computed for both the repulsive
potential (solid curve) and the attractive potential with one bound
s-state (short-dashed curve) defined in the text. The case of the
potential introduced in ref. \protect\cite{zhu93} (with an increase of
3\% of the range of the potential) is also shown (long-dashed line).
}
\label{9}
\end{figure}

In fig. \ref{9} we show the invariant mass spectrum (\ref{invsp}) from a
$^6$He neutron removal reaction. The cases with repulsive s-potential
(solid line) and attractive s-potential with a Pauli forbidden state
(short-dashed line) are shown. The parameters correspond to cases 1 and
7 in table I, respectively. As seen from the figure, both curves are
very similar, and show a peak around $E_{\mbox{\scriptsize core+n}}=0.8$
MeV, that corresponds to the energy of the $p_{3/2}$ resonance.
Another p-resonance is present at 2 MeV in $^5$He. However its width is
much larger than before, and therefore its effect on the spectrum is
much more spreaded out.  To illustrate how the invariant mass spectrum
is sensitive to the resonance energies we also show the case where the
neutron-$^4$He potential introduced in ref.  \cite{zhu93} is used
(long-dashed line). The suggested increase by 3\% of the range of the
neutron-$^4$He potential makes the energy of the $p_{3/2}$ resonance
too small (around 0.3 MeV), giving rise to the pronounced peak in the
invariant mass spectrum at that energy.  The different height of the
curves in fig. \ref{9} comes from the fact that all the three curves are
normalized to 1.

\begin{figure}[t]
\epsfxsize=12cm
\epsfysize=7cm
\epsfbox[550 200 1100 550]{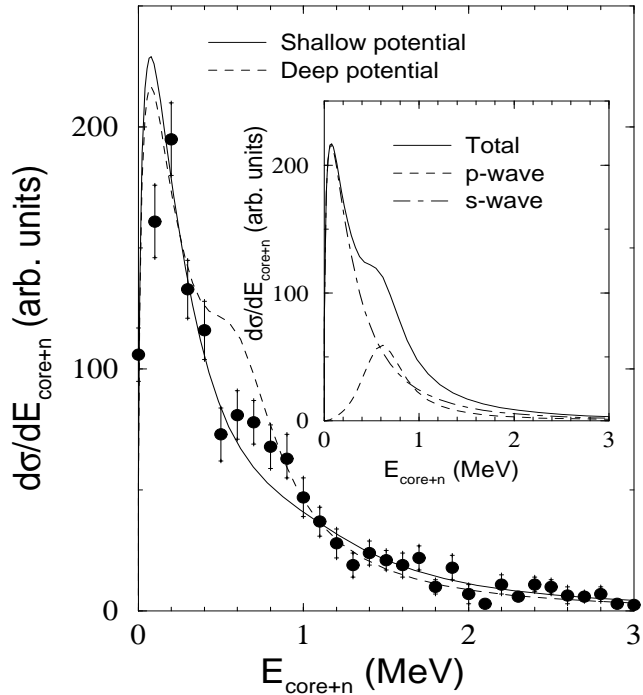}
\vspace{3.1cm}
\caption{\protect\small
The invariant mass spectrum for a $^{11}$Li
neutron removal  reaction computed for both the shallow potential
(solid curve) and the deep potential with one bound s-state
(short-dashed curve) defined in the text. In the inset the case of the
deep potential (solid curves) and the contributions from the s-waves
(dot-dashed line) and the p-waves (dashed line) are shown. The
experimental points for fast $^{11}$Li (280 Mev/u) colliding with a 
carbon target are from \protect\cite{eml96}.
}
\label{10}
\end{figure}

In fig. \ref{10} we show the same spectrum as in fig. \ref{9} from a neutron
removal $^{11}$Li reaction. The cases of the shallow and deep
potentials are shown (cases 1 and 2 in table II). The difference
between these two cases comes from the different energy of the lowest
p-resonance. 

For the deep potential (dashed line) the $p_{1/2}$ resonance is at 0.77
MeV, creating a clear shoulder in the curve at that energy. Note that
the contribution to the wave function from the p-wave (21\%
probability) is given directly by the schematic model. It is determined
by the energies of the virtual $s_{1/2}$ state and the $p_{1/2}$
resonance together with the requirement of fitting the binding energy
and radius of $^{11}$Li. Therefore the shoulder in the distribution is
less pronounced than in calculations where the p-wave content is chosen
to be higher, for instance in \cite{han95}, where a rather arbitrary
50\% contribution for the p-waves is chosen.  For the shallow potential
(solid curve) the large width of the resonance at 1.7 MeV broadens the
corresponding peak at that energy and it disappears completely. For
both the deep and the shallow potential the lowest peak comes from the
low lying virtual s-state in $^{10}$Li. However in this case the
position of the peak is not directly related to the energy of the
virtual state.  A virtual state produces an increase 
\vspace{-10mm}
of the momentum
distributions at zero momentum. In fact, the value of the invariant
mass spectrum divided by $\sqrt{E_{\mbox{\scriptsize core+n}}}$ is not
zero at zero energy, and the lower the virtual s-state the larger the
value at the origin. In the inset of fig. \ref{10} we show the deep potential
case where we have separated the contributions from the s and the
p-waves. It is then clear how the s-wave is responsible for the first
peak, while the shoulder is produced by the $p_{1/2}$ resonance at 0.77
MeV. The experimental points \cite{eml96} correspond to a $^{11}$Li
fragmentation process on carbon. A better agreement between the
computed curves and the experimental data is possible when the spin of
the core is taken into account in the description of $^{11}$Li and
$^{10}$Li.

\section{Summary and conclusions}

We study two prescriptions to take the Pauli principle into account in
the three-body cluster model where more than one cluster contain
nucleons. One is to exclude the Pauli forbidden two-body state from the
active space available for the three-body system, the other is to
construct a two-body potential without the Pauli forbidden state but
with the same low energy properties.

We solve the Faddeev equations in coordinate space by means of
the adiabatic hyperspherical expansion. We calculate the angular
eigenvalues of the Faddeev equations which are closely related to the
effective radial potentials.  This spectrum is computed for the two
different potentials with identical low energy properties and either
without or with one Pauli forbidden bound s-state.  Each bound state
gives rise to a specifically behaving angular eigenvalue. The
technically inexpensive prescription to account for the Pauli principle
is then simply to omit the angular eigenvalue which corresponds to the
Pauli forbidden bound state in the calculations of the three-body
radial wave functions.

We apply these prescriptions to analyses of the high energy
fragmentation reactions of two halo nuclei $^6$He and $^{11}$Li within
a three-body neutron+neutron+core model.  The potentials of the two
prescriptions are adjusted to reproduce the same s- and p-wave low
energy data.

For $^6$He these data are known experimentally and the
interactions are therefore fixed. The potential with forbidden state
then gives correct binding energy and root mean sqare radius for
$^6$He. The equivalent potential without a bound state slightly
underbinds the system. A small reduction of the repulsive core is
needed to obtain the correct binding. This automatically provides
almost the same reasonable root mean square radius.  After this small
adjustment the wave functions for the two nuclei for the two types of
potential are compared and found remarkably similar.

For $^{11}$Li we use the currently accepted experimentla data for the
lowest s- and p- levels in the neutron core subsystem to adjust the
neutron core potentials. With these levels being close to experimental
data the deep potentials provide reasonable binding energy and root
mean square radius. The equivalent potential without a bound state
slightly overbinds the system. After a small attenuation of the
position of the neutron core $p_{3/2}$ level this prescription also
provides the correct binding and size of the system.

In general the two prescriptions without any fine tuning provide very
close but still distinguishable ground state properties of the three-body 
system. After small adjustments the properties become remarkably
similar.

We then compute neutron and core momentum distributions in nuclear
break-up reactions of these halo nuclei. We use the sudden
approximation and include final state interactions, which are crucial
for the neutron distributions as the full width at half maximum for the
neutrons from $^6$He is reduced by a factor of more than 2 from the value
corresponding to omitting the final state interaction.

The two prescriptions give almost identical results and the measured
momentum distributions are rather accurately reproduced. These
computations are carried out in a consistent model where the same
two-body potential is used both for the initial three-body halo
structure and for the final state interaction after the break-up
process.

The final state invariant mass spectra for $^5$He and $^{10}$Li are
computed and their features discussed. A virtual s-state shows up as a
pronounced peak close to zero energy. This peak arises from the phase
espace factor which vanishes at zero energy, and its position is
unrelated with the energy of the virtual state. Higher angular momenta
show up as distinct peaks at the resonance energy. These spectra are
then sensitive to the continuum structure of the two-body system.

In conclusion the two prescriptions to account for the Pauli principle
work remarkably well for the two test examples $^6$He and
$^{11}$Li. In both cases the lowest Pauli forbidden s-state is
occupied by nucleons in the core nucleus. However, the valence
neutrons occupy predominantly neutron-core relative p-states and
s-states, respectively for $^6$He and $^{11}$Li. The wave functions
are further tested in connection with our fragmentation model and the
measured momentum distributions are nicely reproduced. For $^6$He
these quantities are computed for the first time whereas $^{11}$Li was
investigated earlier in a more sophisticated model.  Although fairly
realistic, we used here only a schematic model, because the main
purpose was to demonstrate the reliability to account for the Pauli
pinciple.  In any case the results provide additional support both for
the prescriptions and for the fragmentation model.

\paragraph*{Acknowledgments.} 
We want to thank B. Jonson, K. Riisager, and H. Emling for useful
discussions and for making the latest unpublished experimental data
available.  One of us (E.G.) acknowledges support from the European
Union through the Human Capital and Mobility program contract nr.
ERBCHBGCT930320.

\end{document}